\documentstyle[psfig]{mn}

\input{epsf}

\begin{document}

\title[The galaxy luminosity function around groups] 
{The galaxy luminosity function around groups.}
\author[Gonz\'alez et al.]{
\parbox[t]{\textwidth}{
Gonz\'alez, R. E., Padilla, N.D., Galaz, G., \& Infante, L.}
\vspace*{6pt} \\ 
Departamento de Astronom\'\i a y Astrof\'\i sica, Pontificia
Universidad Cat\'olica de Chile, V. Mackenna 4860, Santiago 22, Chile.\\
}
\maketitle

\begin{abstract}
We present a study on the variations of the luminosity function of galaxies around
clusters in a numerical simulation with semi-analytic galaxies, attempting to 
detect these variations in the 2dF Galaxy Redshift Survey.
We subdivide
the simulation box in equal-density regions around clusters, 
which we assume can be achieved by selecting objects
at a given normalised distance ($r/r_{rms}$, where $r_{rms}$ is an
estimate of the halo radius) from the group centre.
The semi-analytic model predicts important variations in the luminosity
function out to $r/r_{rms}\simeq5$.  In brief,
variations in the mass function of haloes around clusters (large dark-matter
haloes with $M>10^{12}$h$^{-1}$M$_{\sun}$) lead to 
cluster central regions that present a high abundance of bright galaxies
(high $M^*$ values) as well as low luminosity galaxies (high $\alpha$); at $r/r_{rms}\simeq 3$
there is a lack of bright galaxies, which shows the depletion of galaxies
in the regions surrounding clusters (minimum in $M^*$ and $\alpha$),
and a tendency to constant luminosity function parameters at larger
cluster-centric distances. 
We take into account the observational biases present in the real
data by reproducing the peculiar velocity effect on the redshifts of
galaxies in the simulation box, and also by producing mock catalogues.
We find that excluding from the analysis galaxies wich in projection are close
to the centres of the groups provides results that are qualitatively
consistent with the full simulation box results.
When we apply this method to mock catalogues of the 
2dF Galaxy Redshift Survey (2dFGRS) and the 2PIGG catalogue of groups, 
we find that the variations in the luminosity function are almost completely erased
by the finger-of-god effect; only a lack of bright galaxies at $r/r_{rms}\simeq 3$ 
can be marginally detected in the mock catalogues. 
The results from the real 2dFGRS data shows a more clear detection of a dip
in $M^*$ and $\alpha$ for $r/r_{rms}=3$, consistent with the semi-analytic
predictions.  
%The results from the full 2dFGRS galaxy sample provides
%a new measurement of the galaxy luminosity function, which can be well described by
%a Schechter function with parameters $\alpha=-1.26 \pm 0.03$, $M^*=-19.84 \pm 0.11$,
%and $\phi^*=0.0159 \pm 0.0030$.
\end{abstract}


\section{Introduction}
\label{sec:intro}

Understanding the way in which galaxies form is a key challenge
facing cosmologists today.  This can be approached in different ways,
of which the most important can be
 the study of the clustering of galaxies and the bound systems
defined by their spatial distribution, and the
statistical measures of the properties of galaxies and how these
vary with their environment.

There have been several attempts to 
characterise galaxy properties as a function
of local density \cite{bromley,garilli,ramella,croton}.
In particular, studies of the variation of the luminosity function
include the luminosity function of wall and void galaxies in the
2dFGRS \cite{croton}, in the SDSS \cite{hoyle},
and theoretical variations in the luminosity function as a function of
local galaxy density \cite{mo}.

In this paper we concentrate on the study of the
variations of the luminosity function with local density, which
we characterise by the distance to the nearest group of galaxies, normalised
by the group radius.
We approach this problem by first studying
numerical simulations with semi-analytic galaxies, to later
corroborate with mock 2dFGRS catalogues the possibility of applying
the method to the real 2dFGRS galaxies, thus
providing a new test of the semi-analytic model of galaxy formation.  

It should be noted that according to the present understanding
of the galaxy formation process, galaxies form exclusively
in dark-matter haloes \cite{kauff,cole,somer,naga}.  Therefore, our analysis 
does not refer to groups as any dark matter halo, but instead
to associations of at least two galaxies with
$B_J<-16.5$.  This, according to recent measurements of the halo 
occupation 
number statistics \cite[in SDSS]{abaz2}, can only be found in haloes
with $M>10^{11}$M$_{\odot}$.
Other studies \cite{lemson} shows a dependence of the halo mass function on 
the environment.  Since the properties of the semi-analytic galaxies depend
only on their host halo mass (\footnote{The only input to the semi-analytic model 
coming from the numerical simulation is the halo mass. Therefore,
any influence from large scale (i.e. the environment external to the halo in
the simulation) has no direct influence on the resulting semi-analytic galaxies. 
On the other hand, the physics implemented by the semi-analytic model can have
a much more noticeable impact on the properties of semi-analytic galaxies; in
this work, the physics governing the evolution of galaxies is held fixed and
correspond to the Benson et al. (2003) model.}), 
we can also expect a dependence of the luminosity function 
on the local environment.

This paper is organised as follows.
We start by describing the numerical simulation and semi-analytic
galaxies used in this work in Section \ref{sec:sim}, where we also
describe the selection of haloes used to characterise the
local galaxy density in the simulation box.  In Section \ref{sec:method} we
describe the method and algorithms used to determine the luminosity
function corresponding to the different semi-analytic galaxy 
environments.
We present the results from the numerical simulation galaxies
in Section \ref{sec:simres}, and study the possibility of detecting
such results in the 2dFGRS by applying the method to mock 2dFGRS
catalogues in Section \ref{sec:mocks}.  In Section \ref{sec:results}
we show the results from the real 2dFGRS catalogue, and finally
in Section \ref{sec:conc} we conclude our work with a summary
of our main conclusions.

\section{The numerical simulation}
\label{sec:sim}

The numerical simulation used in this work,
kindly provided by the Cosmology group at Durham group,
comprises data
on dark matter particles and on GALFORM semi-analytic galaxies.
The numerical
simulation box is $250$h$^{-1}$Mpc a side, and contains
$125,000,000$ particles.  The process by which the distribution
of dark matter particles was populated by galaxies can be
found in Cole et al. (2000), and can be summarised as follows.
After haloes of at least $10$ dark matter particles are identified
in the simulation using a Friends-of-friends algorithm, 
the GALFORM algorithm produces a Monte-Carlo merger history for each halo
which defines the galaxies that populate it. We emphasize that
the properties of these galaxies depend only on the halo mass, and not on the
environment in which each halo resides.
These galaxies
are characterised by magnitudes in several bands (of which
we only make use of the $B_J$ band), and stellar formation
rate parameters, plus spatial positions and peculiar velocities.
The total number of galaxies in the simulation box is approximately
$1.8$ million.  Due to the dynamical limitations of the numerical
simulation, the galaxy sample is limited by an absolute magnitude
limit of $B_J<-16.5$. 
\section{Method}
\label{sec:method}

\begin{figure}
{\epsfxsize=11.truecm 
\epsfbox[20 160 775 745]{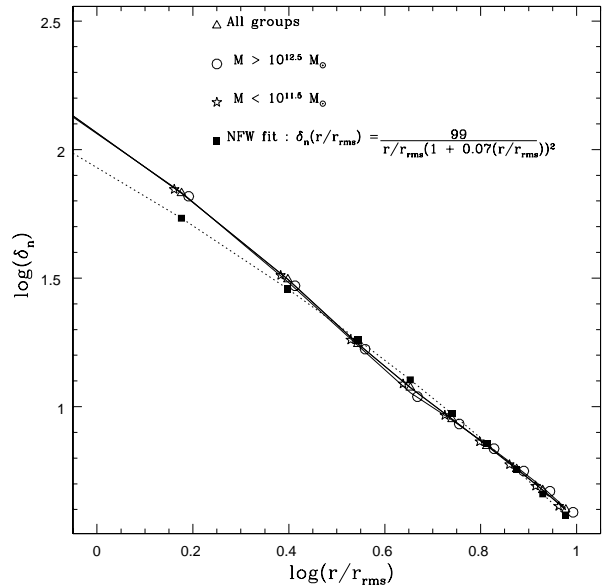}}
\caption{
Comparison between the density profiles of galaxies and mass around haloes
in the numerical simulation cube.
}
\label{fig:dens}
\end{figure}

In this paper we characterise the density at which a galaxy
resides by locating the dark-matter halo that is closest
to the galaxy position in units of halo/group radius, denoted by
$r_{rms}$, calculated as half the mean projected distance between
pairs of galaxies in a group.  
We use a normalised distance ($r/r_{rms}$)
in order to take into account the different halo sizes. This is motivated by
the clear indications from numerical simulations that the galaxy
density around dark-matter haloes depends on the normalised
distance and on the halo concentration parameter \cite[NFW]{nfw}.  
Figure \ref{fig:dens} shows the density profile
around haloes for the dark-matter (dotted line) using NFW profile, and for galaxies 
in halos with different mass ranges in the simulation (solid lines). 
The galaxy density profile shows no important variation with the halo mass.   
As can be seen, even though the halo
concentrations vary between the different samples of haloes 
shown in the figure, there is little difference between the
dark-matter and galaxy density profiles, even for $\log(r/r_{rms})<0.3$,
the dark-matter density is only lower than the galaxy density by about a factor of $0.9$.
This fundamental result supports our hypothesis that a sample of galaxies lying within a given 
range of normalised distances to halo centres are characterised by the same local 
galaxy or dark-matter density.

\begin{figure}
{\epsfxsize=11.truecm 
\epsfbox[0 -100 2900 1900]{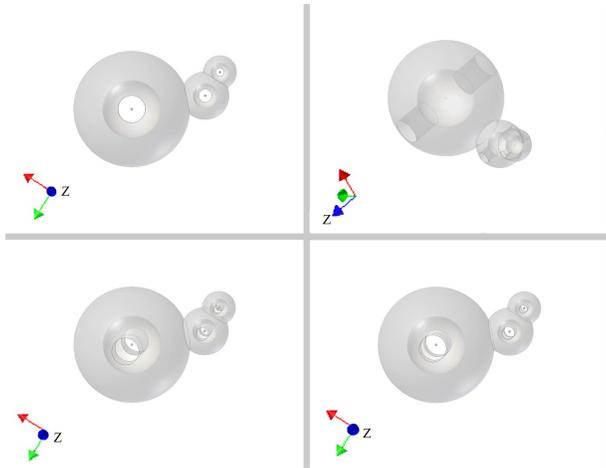}}
\caption{
Schematic representation of the geometry of the problem.  
The subsample of galaxies selected according to $1<r/r_{rms}<2$ fills the spherical
volumes depicted in this figure.   
The cylindrical volumes are oriented in the direction of the line of 
sight (z-axis), and are removed from our analysis when effects of peculiar velocities
need to be avoided.
}
\label{fig:geometry}
\end{figure}

The procedure to define our semi-analytic galaxy
subsamples can be described as follows: 1. Calculate the distance
from each galaxy to the nearest haloes. 2. Assign this galaxy
to the halo that presents the shortest normalised distance $r/r_{rms}$ to the
galaxy. 3. Divide the galaxy sample according to the normalised
distance to the closest halo.  This process produces galaxy subsamples
that are characterised by a complicated spatial distribution
which is depicted in Figure \ref{fig:geometry} from different viewpoints
(Top left: Front view, Top right: Side-view, Lower panels, angle views); 
in this example we have three
shells of equal density for three groups of galaxies (each sphere corresponds
to a different group).

As we are interested in assessing the possibility of applying
this method to real data, the different observational biases present in
real data samples have to be taken into account as well.
One of the effects that will prove particularly important
for our work is that of peculiar velocities in redshift surveys, which
causes the well known finger-of-god effect around groups and clusters of
galaxies.  This effect makes a galaxy that is sitting in the middle
of the cluster potential well appear spuriously far away from
the cluster centre.  This will lead us to induce that
this galaxy belongs to a subsample characterised by a low density
environment due to its spuriously large separation from the halo 
centre and will introduce a systematic error in our measurements.  In order
to avoid this problem, we simply remove a cylinder centered on the
halo centre and aligned with the direction of the line of sight.
In the case of the simulation box, we take this direction to
be the z-axis. Note that in Figure \ref{fig:geometry} we have removed 
these cylinders from the spheres surrounding each group.

The galaxy subsample selection criteria and the removal of
a cylinder shaped volume to avoid the effects of peculiar velocities
makes it difficult to estimate the normalization of the luminosity
function.  Therefore, in cases when an estimate of the volume is needed we also 
produce a catalogue of random
positions which populate the same regions occupied by the
galaxy subsample of interest.  Such random catalogues are produced using 
the same halo centres in the numerical simulation and following the steps used for
defining the galaxy subsamples, but
this time using the random particles instead of semi-analytic galaxies.

The luminosity functions (LF hereafter) are calculated using the standard STY (Sandage, Tammann
\& Yahill, 1979), Step Wise Maximum Likelihood (SWML, Efstathiou, Ellis, \& Peterson, 1988)
and $1/V_{MAX}$ (VMAX, 
Schmidt, 1968, Huchra \& Sargent, 1973) methods. As Willmer (1997) provides
a detailed review on the three different methods, we only present their most
relevant characteristics. The VMAX method is the most popular of the three,
and weights each galaxy by the volume corresponding to the maximum distance 
out to which it would remain below the magnitude limit of the survey.
This method also provides the normalization 
of the LF, but it does not take into account the fact that galaxies present
a clustered distribution.  Another setback of the VMAX method is that it
requires the counts of galaxies to be binned in magnitude,  
which makes data of individual galaxies to be lost. The 
STY method is a maximum likelihood estimator, which has the advantage of
calculating a likelihood for each individual galaxy, as well as being an unbiased 
estimator of the LF for a clustered 
spatial distribution of galaxies. One drawback is that this method does 
not provide the normalization of the LF and assumes a fixed functional form for
the shape of the LF, usually a 
Schechter function (Schechter, 1976).  The 
SWML method is similar to the STY method, in that it is also a maximum likelihood
method, but it does not 
assume a specific shape of the LF since it uses galaxies binned in consecutive absolute 
magnitude intervals.  In the case of the STY and SWML methods, the 
normalization is calculated using the estimators given by \cite{dh}.

We will quantify the luminosity function by quoting the values
of the parameters $M^*$ and $\alpha$ in the Schechter formula:
\begin{eqnarray}
\qquad \qquad \phi (M)dM = 0{.}4 \  A \ln 10 \phi ^*e^{-X}X^{\alpha + 1}dM, \nonumber
\end{eqnarray}

\noindent
where $ X = 10^{0{.}4(M^* - M)}$, and the parameter $M^*$ refers
to the characteristic luminosity of the sample of galaxies;
$\alpha$ is the faint-end slope and indicates the relative importance of a population
of low luminosity galaxies.  For instance, a large, negative
value of $\alpha=-1.7$ indicates a large population of
dwarf galaxies; $\phi^*$ is the normalization of the LF which we choose not to calculate 
since this parameter traces the density profile around groups of galaxies or
dark-matter haloes.

\section{Numerical simulation results}
\label{sec:simres}

The mass function of haloes in the numerical simulation
is well described by the Jenkins et al. (2001) formula, which
is characterised by a large population of low mass haloes,
and a few, rare, high mass objects with $M>10^{15}$M$_{\odot}$.
Observational catalogues face little difficulties in detecting
the high mass objects, but cannot observe galaxies fainter than the apparent 
magnitude of the survey, which translates into an approximate
lower mass cut-off in the catalogue.
%miss most of the objects below a lower
%mass limit.  
For example, in the 2dFGRS Percolation Inferred Galaxy
Group catalogue (2PIGG), this is
$M_{low} \simeq 10^{11}$M$_{\odot}$ (Eke et al. 2004).

\begin{figure}
{\epsfxsize=8.truecm 
\epsfbox[20 160 575 775]{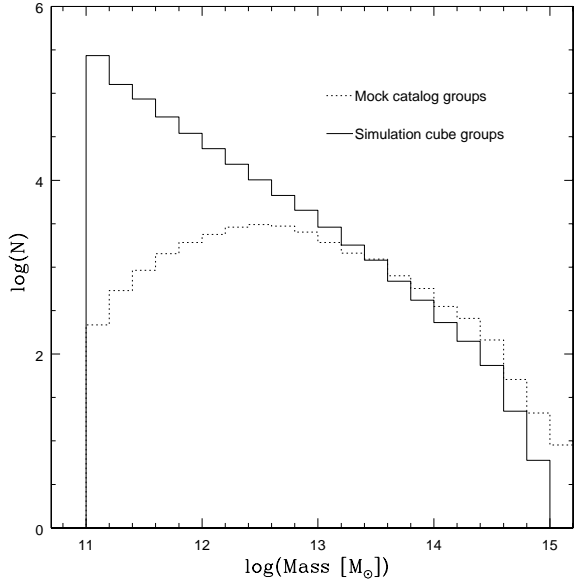}}
\caption{
Selection of haloes from the numerical simulation box so as
to match the mass distribution of the mock catalogue.
}
\label{fig:selgroup}
\end{figure}

In order to illustrate what would be found in the dependence
of the luminosity function with normalised distance to the halo
centres, we first select from the simulation, haloes with 
a distribution of masses resembling that of
an observational sample of groups.  In particular, we use the
distribution of masses in the full 2PIGG presented by Eke et al. (2004). 
In Figure \ref{fig:selgroup} we show the distribution of masses in
the simulation and in a 2dFGRS mock catalogue.  We then select haloes
from the simulation so that they show the same distribution of masses
as the 2PIGG.
As can be seen, this is very similar to placing a lower
limit on the halo mass of $M>10^{11}$M$_{\odot}$.

\begin{figure}
{\epsfxsize=8.truecm 
\epsfbox[20 160 575 775]{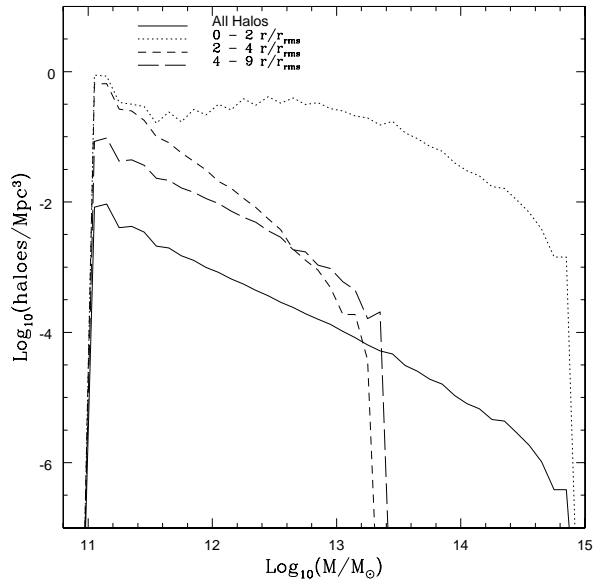}}
\caption{
Mass function of haloes at different distances
from ``centre haloes" (see text).  Distance
ranges are shown in the key.
}
\label{fig:massf}
\end{figure}

It should be noted that as the semi-analytic model for galaxy
formation only takes into account the mass of the dark matter
halo when populating it with galaxies, the dependence
of the luminosity function with normalised distance responds
to variations in the mass function of low mass haloes around
larger dark-matter haloes (in order to improve clarity, we 
will refer to the latter as simulated clusters).
Figure \ref{fig:massf} shows the
variations in halo mass function as the normalised distance
($r/r_{rms}$) to a 
nearby simulated cluster increases.
The differences in amplitude
respond mainly to the variation in density around
simulated clusters; for instance, in the cluster central regions,
the density is about $2$ orders of magnitude larger
than the mean in the simulation, thus explaining the
$2$ order of magnitude difference between the dotted
and solid lines.  In this Figure there is a striking
lack of high mass haloes in the mass function of haloes in
the simulated cluster outer regions compared to the inner haloes
at $r/r_{group}<2$; this is caused by the selection of
clusters in the simulation, which contains all the high
mass haloes with masses $M>10^{13}$h$^{-1}$M$_{\odot}$.  
The resulting mass functions show
larger abundances of high mass haloes in the cluster centres, a
slight decrease in high mass haloes for the
intermediate distances, and again an increase for the larger
distances from the simulated cluster centres. 
On the low mass end of the mass function, 
it can be seen that the inner haloes present the largest population,
whereas this diminishes as we go towards haloes in regions
further away from the cluster centres.

Variations in the mass function of low mass haloes around simulated clusters
may be biased in the sense that simulations show that the radial
distribution of sub-haloes within their parent haloes is substantially less
centrally concentrated than that of the dark matter \cite{gao,ruso}.  
This is not a setback to our procedures since we only argue that
the normalised distance between haloes and clusters provides a reasonable
proxy for local density.  Furthermore, we have shown that semi-analytic
galaxies trace the same density profile regardless of the mass of the
central cluster.

Given that the properties of semi-analytic galaxies depend
only on its host halo mass,
the variations in the mass function of haloes on normalised distance
to simulated clusters will imprint variations in the LF.
For instance, the haloes in the outermost regions contain larger abundance 
at $M \simeq 10^{13}$M$_{\odot}$ which contributes satellites galaxies which
may increase the value of $\alpha$. The higher upper limit in halo mass will also
induce a change in $M^*$ since larger haloes tend to host the 
brightest galaxies.

In order to study the variations of the LF in the simulation,
we proceed to rank the semi-analytic galaxies according
to their normalised distance to the closest halo centre, and
produce several subsamples at $0<r/r_{rms}<1$, $1<r/r_{rms}<2$,
and so forth.  In a first instance, we use the real-space 
galaxy positions.  Later on, we apply redshift-space
distortions by displacing the galaxy z-coordinate positions by an
amount equal to the z-component of their peculiar velocities 
(in units of h$^{-1}$Mpc).  

\begin{figure}
\begin{picture}(230,450)
\put(0,0){\psfig{file=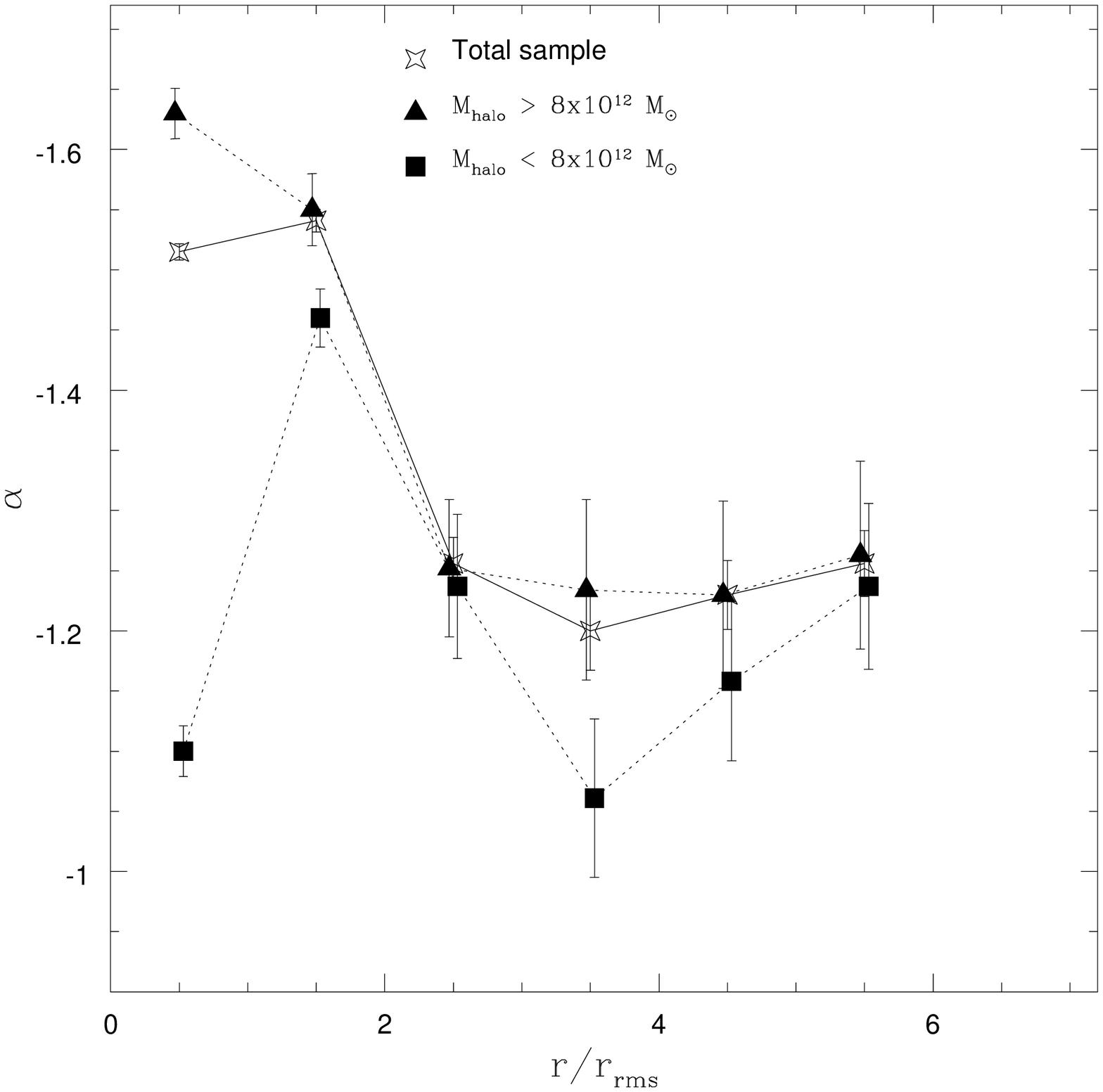,width=8.cm}}
\put(0,215){\psfig{file=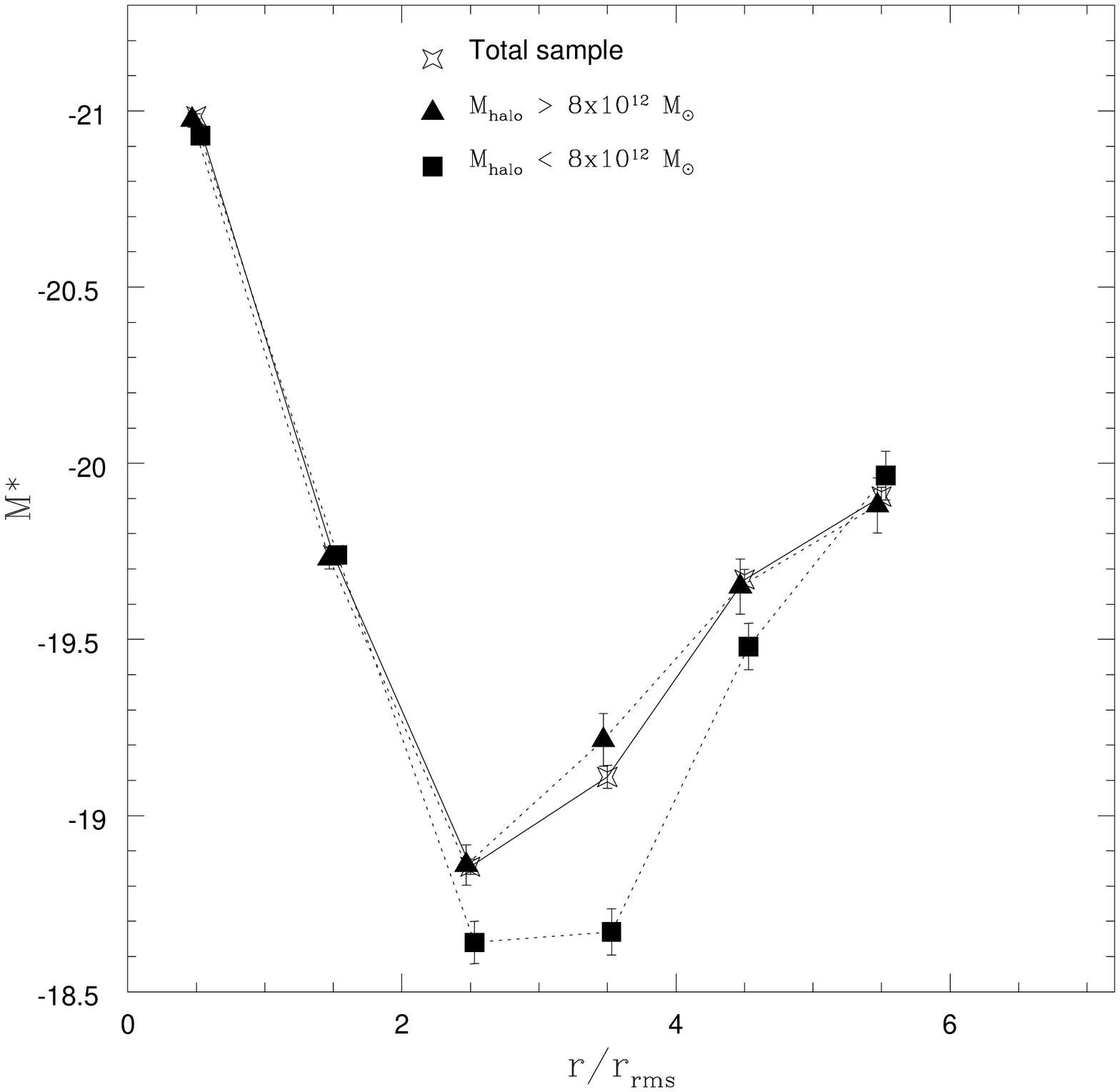,width=8.cm}}
\end{picture}
\caption{
Top panel:
Variation of the characteristic absolute magnitude $M^*$ parameter 
as a function of normalised distance
to the group center.  Different lines correspond to different 
ranges in simulated cluster masses (indicated in the key).
Bottom panel:
Same as top panel but for the faint-end slope of the LF $\alpha$.
}
\label{fig:cube}
\end{figure}

The results in real-space are displayed in solid lines in Figure \ref{fig:cube}, where
the top panel shows the dependence of $M^*$ on $r/r_{rms}$,
and the bottom panel, the dependence of the parameter $\alpha$.
Error bars correspond to $1$-$\sigma$ errors
calculated using the Jacknife method.  All errors in this work are
calculated using this method, a choice that
is supported by numerous works reporting that Jacknife errors are comparable
to the dispersion of results from large numbers of realistic mock
catalogues (See for instance, Padilla et al., 2005).
As can be seen, the dependence of $M^*$ indicates a
population of very bright galaxies near the simulated cluster
centres, which dims significantly as we approach a
normalised distance of $r/r_{rms}\simeq3$.  Interestingly,
at $r/r_{rms}\simeq 5$, the characteristic luminosity
increases to reach a constant value of $M^*=-20$, which is consistent
with the value of $M^*$ characterising the full sample of semi-analytic
galaxies in the simulation.
This in perfect agreement with what was found for the mass function of haloes in the
same normalised distance ranges.   

Qualitatively, a very similar behaviour can be seen for the parameter
$\alpha$, which is very negative at the centre of the groups,
indicating a large population of low luminous galaxies or semi-analytic satellites
(low and high mass haloes respectively).
At $r/r_{rms}\simeq 3$, $\alpha$ reaches its highest value (or flattest slope)
$\alpha\simeq -1$, and settles at $\alpha\simeq-1.25$ at
larger normalised separations.  The full sample of galaxies
in the simulation shows $\alpha=-1.37$; therefore the galaxies
at large separations from groups 
show a comparable $M^*$ to the full sample of galaxies,
but a comparatively low population of dwarf galaxies.

These findings encode a wealth of information regarding
the galaxy formation process and possible further evolution. Even though the semi-analytic
model for galaxy formation is a relatively simple one,
in the sense that it only requires the mass of a host halo in order
to populate it with galaxies, it manages to produce answers
that can be reconciled with several physical processes that
affect the evolution of galaxies.  For instance, the dependence
of the parameters $M^*$ and $\alpha$ can be understood as follows.

The high luminosity galaxies in the centres of groups along with
a large population of dwarfs is a reasonably well documented
fact observed in redshift surveys \cite{blanton}
and individual clusters \cite{rob,valotto}, and is
thought to be a consequence of subsequent mergers of haloes. This
preferentially takes place in the centres of groups and clusters of 
galaxies (massive haloes), where the galaxy density is high.  The large population of
dwarf galaxies is thought to be produced by the close interaction of haloes, tidal
stripping \cite{stadel} or other processes considered in the semi-analytic model.  
Note that since this simulation only takes into account dark-matter
interactions, no gas processes are taken into account.

A new interesting result that can be appreciated in Figure \ref{fig:cube}
is how extended is the influence of simulated clusters over their
surrounding medium.  As can be seen, the population of low luminosity galaxies (i.e.
low mass haloes) is slightly reduced at
distances $r/r_{rms}\simeq3$ (faint $M^*$ and flat $\alpha$) with respect to the field population,
marking the radius of influence of clusters.
The mechanisms that produce such effects
take place in the cluster centres, so this loss of galaxies may be the consequence
of galaxy/halo passages through the near centres of clusters.  There
is also a clear depletion of low luminosity galaxies (low absolute
values of $\alpha$), which can also be understood in terms of passages
of low mass haloes (dwarfs) through the group centres of which a fraction is cannibalised
by larger haloes.  

Finally, at separations $r/r_{rms}\simeq6$, the behaviour of
both $M^*$ and $\alpha$ tends to constant luminosity
function parameters for larger separations.
Notice that the values of the field LF parameters in the simulation
depend on input values that have not been tuned to reproduce exact
LF parameters of 2dFGRS data.

We also check whether the variations in luminosity 
function depend on the mass of the nearby simulated clusters. Figure
\ref{fig:cube} shows in solid squares and triangles, the
resulting variations in $M^*$ and $\alpha$ with 
$r/r_{rms}$ when the mass of the centre haloes varies (solid squares
for low mass haloes, solid triangles for high mass haloes).
As can be seen, the characteristic luminosity does not
change significantly in the central or more external regions,
being the values of $M^*$ consistent when taking as centres
either low, high or all the haloes selected.  The major
difference can be seen at intermediate distances, $r/r_{rms}\simeq 3$,
where low mass haloes are surrounded by lower luminosity galaxies
than higher mass haloes (by almost half a magnitude).  Regarding the distance
at which $M^*$ tends to a roughly constant value,
it can be argued that lower mass haloes influence a larger volume in terms
of their radius, $r_{rms}$, since $M^*$ remains at low luminosity values
out to $r/r_{rms}\simeq3.5$.
The variation in the luminosity function slope $\alpha$ shows that
at intermediate distances, low mass haloes contain less low luminosity
galaxy neighbours than high mass haloes.  On the other hand, at low
$r/r_{rms}<1.5$, the population of faint galaxies decreases dramatically
in low mass haloes.  In this case the low mass halo sample contains objects
with $M<8 \times 10^{12}$h$^{-1}$M$_{\odot}$, which for the magnitude limit
of the semi-analytic galaxies in this simulation, corresponds to haloes that
preferentially contain $2$ or less galaxies,  
whereas large mass haloes contain large amounts of satellite galaxies which
account for this difference.

\section{Testing the applicability of the method to observational data}
\label{sec:mocks}

\begin{figure}
\begin{picture}(230,450)
\put(0,0){\psfig{file=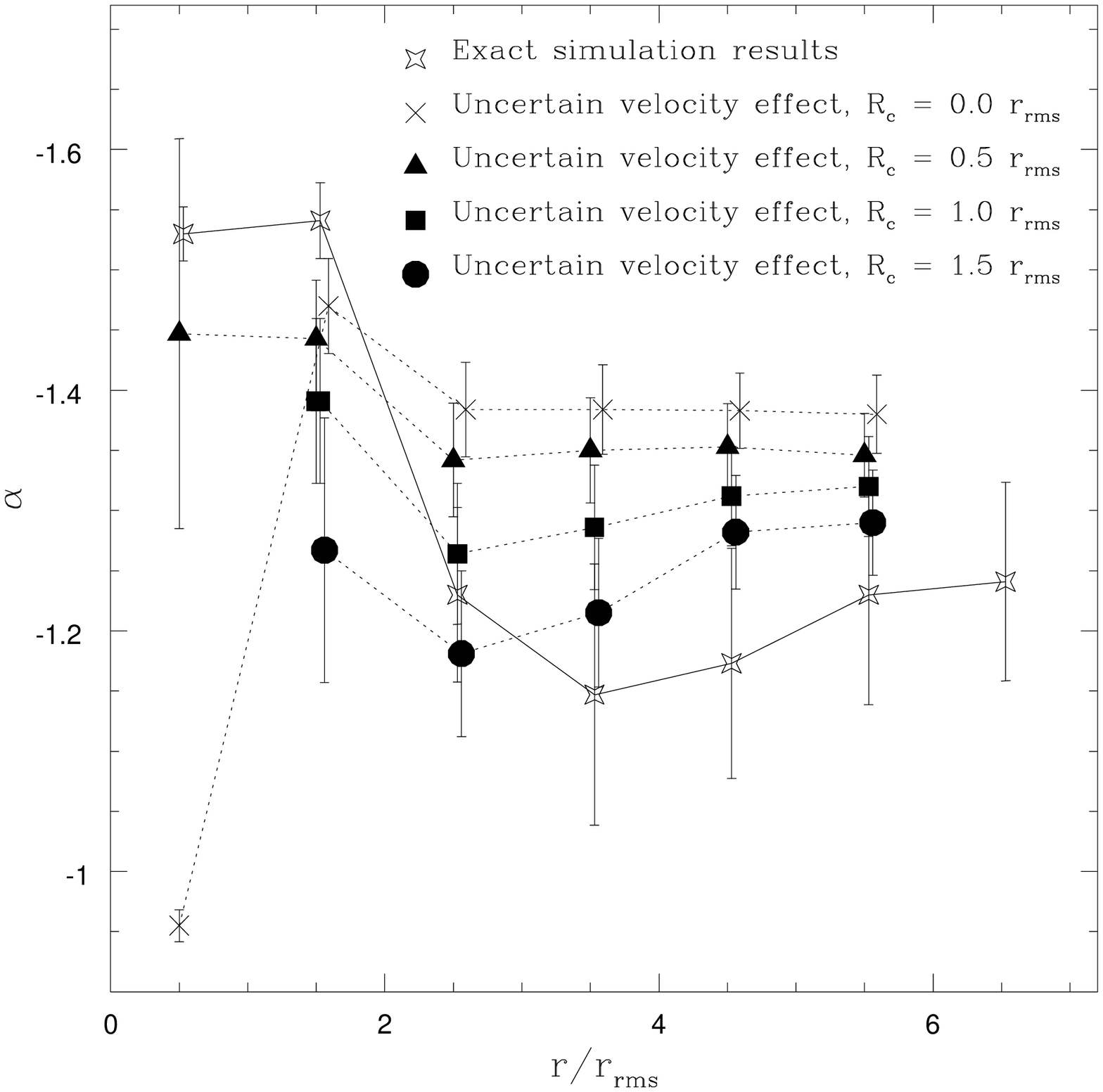,width=8.cm}}
\put(0,215){\psfig{file=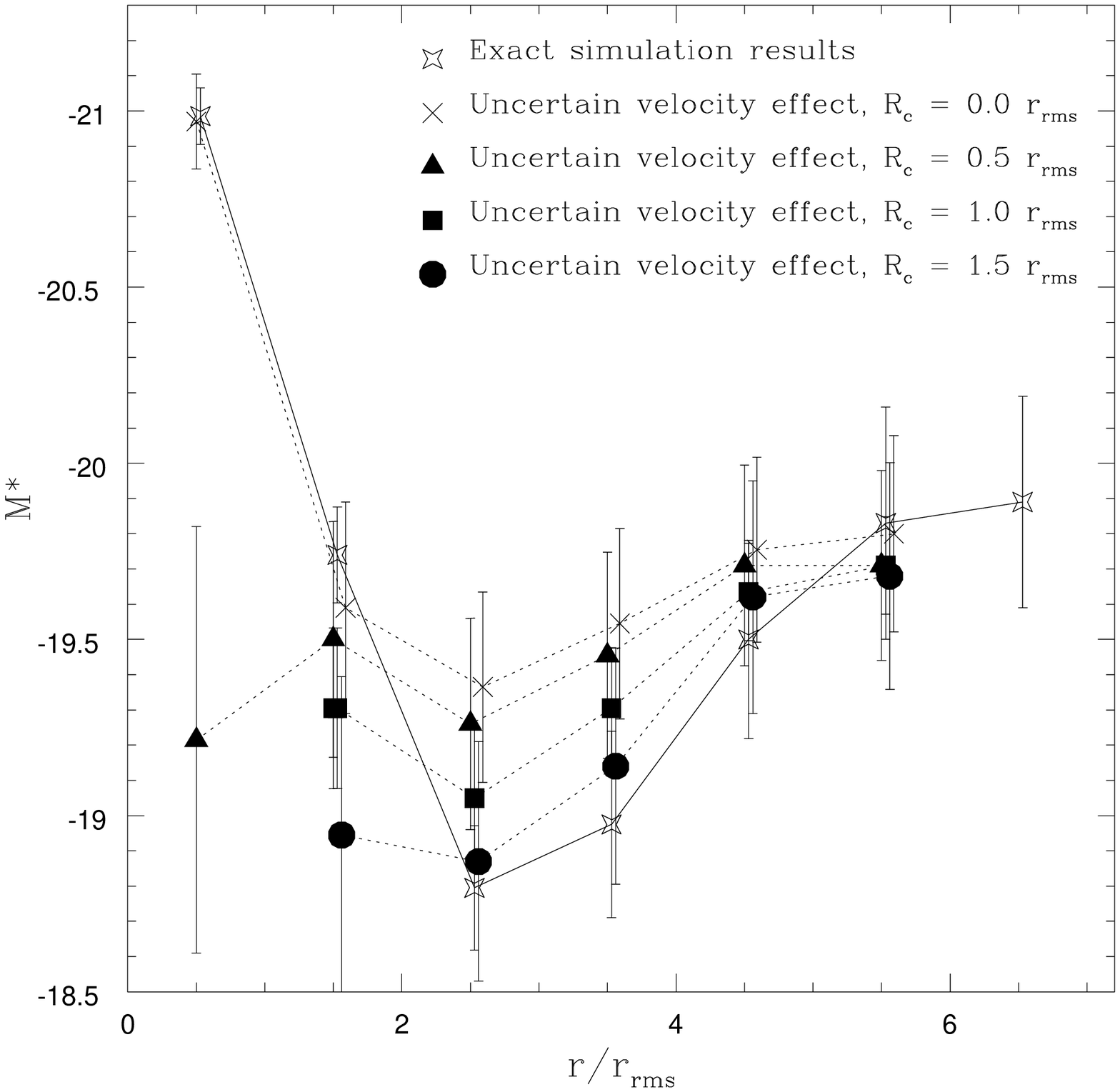,width=8.cm}}
\end{picture}
\caption{
Top panel:
Variation of the characteristic absolute magnitude $M^*$ as a function of normalised distance
to the group center.  Different symbols correspond to different
radii for the cylindrical volume that is removed from the analysis in order
to avoid contamination of the clusters inner galaxies on outer shells
due to the velocity dispersion of the clusters.
Bottom panel:
Same as top panel but for the faint-end slope, $\alpha$.
}
\label{fig:parcube}
\end{figure}

As a first check, we repeat the study performed in the previous
Section to the full simulation box data, but taking the galaxy
positions in redshift-space.  In this case, as can be seen in figure
\ref{fig:parcube} shown by the crosses, the effect of peculiar velocities smears the
relation seen in the cube in real-space, to a degree that makes
it difficult to arrive at any conclusions regarding a dependence
of the luminosity function parameters on normalised distance.  
This is produced by the high velocity dispersions that characterise
the centres of galaxy groups and clusters, which produce the
finger-of-god effect in redshift surveys, and that spuriously
displaces galaxies corresponding to the centres of groups to
the group outer regions, thus contaminating the relation between
$M^*$ and $\alpha$ with $r/r_{rms}$. This is a very important effect that needs
to be taken into account when analysing redshift space data.

In order to correct for this effect, we proceed to remove
from our analysis all those galaxies that lie within a 
cylindrical volume centred on the group centres and aligned
with the direction over which the galaxy positions were
displaced by their peculiar velocities (the z-axis in this
case).  We experiment with different cylinder radii and show
different symbols in Figure \ref{fig:parcube} (see the key).  As can be seen,
this technique is able to partially recover the qualitative
shape of the dependence of both $M^*$ and $\alpha$ with
$r/r_{rms}$ found using the full simulation box in
real-space. Notice that as we increase the cylinder radii to
reduce the peculiar velocity effect, we lose data of central galaxies in groups,
so we test with different cylinder radii.  We select an
optimal value for this parameter
such that it minimises the effects from redshift-space distortions
and low-number statistics, $r_c=0.5r_{rms}$.
This is the value of $r_c$ that we will adopt from now in the analysis of both mock
and real catalogues, unless otherwise stated.

\subsection{The 2dFGRS galaxy and group catalogues}

Before going any further, we describe the characteristics of the observational data.
The 2dFGRS survey \cite[\& 2003]{coll} 
comprises a total of $\sim 230,000$ low
resolution spectra of galaxies 
distributed in two regions near the North and South
galactic poles (NGP and SGP regions respectively)
covering a total of about $2000$ $deg^2$ on the sky.  The
apparent magnitude limit of the survey is of the order
of $b_J<19.4$ with plate-to-plate variations
on the sky of about $0.2$ magnitudes.  The parent
imaging 2dFGRS catalogue is the Automated Plate
Machine survey, APM \cite{madd}. The survey presents
a complicated angular completeness mask which has also been 
made available by the 2dFGRS Team and which we use in all 
of our following analyses.
When measuring luminosity functions using 2dFGRS data, we pay
special atention to remove angular incompleteness effects by
weighting each galaxy by the inverse of the incompleteness factor
at its position (this information is part of the public 2dFGRS data).
This incomplentness factor includes effects from fiber collisions,
and therefore our results are expected to be free from such effects.

The galaxy groups to be used in our analysis are taken
from the 2PIGG catalogue \cite{eke}, which
contains groups of galaxies identified in the 2dFGRS with
at least 2 members.  There is a total of $30,000$
groups in the 2PIGG catalogue.  The advantage
in using this particular sample of groups relies
in the careful calibration of the friends-of-friends
(FOF) method in Eke et al. (2004) using detailed
mock 2dFGRS catalogues.
The distribution of group $r_{rms}$ radii show
only a few hundred groups with $r_{rms}<30$ arcseconds, with
a typical $r_{rms}=4$ arcminutes.  This
indicates that there is little effect on our results
from this problem, which we also correct for by applying a
completeness weight.  At any rate, our conclusions from the 2dFGRS
are only relevant at distances $r>2r_{rms}$ from the group centres;
a study of the group central regions would need a more careful treatment
of this problem.

We will now test whether we can use the 2dFGRS groups and galaxies to detect the variations
in the LF of Section 4 using mock catalogues.

\subsection{Mock 2dFGRS galaxy and group catalogues}

We use the semi-analytic galaxies in the numerical
simulation to produce mock 2dFGRS and 2PIGG catalogues.
We place an observer at a random
position in the simulation box, then simulate
the observing biases including variable magnitude
limits, angular incompleteness masks, and $(k+e)(z)$
corrections from Norberg et al. (2002).
The final mock 2dFGRS catalogue contains comparable
numbers of members in the NGP and SGP areas.

We then feed the mock 2dFGRS catalogue into the FOF
algorithm kindly provided by Vincent Eke, and produce
a mock 2PIGG catalogue.  One of the advantages of the
mock 2PIGG catalogue over the real 2PIGG is that
we also know the true mass of each galaxy group.  The
true mass of a group is that of the halo in the 
numerical simulation box corresponding to the galaxy
group in the mock catalogue.

Therefore, these mock catalogues match as closely as possible
the properties of the observational dataset,
and will make it possible for us to assess whether
the properties of the luminosity function of galaxies
at different normalised distances from galaxy groups can
be measured with the real 2dFGRS dataset.  

%Aqui talves seria clarificador volver a explicar que r_halo = r_rms

\subsection{Application to mock catalogues}

\begin{figure}
{\epsfxsize=11.truecm 
\epsfbox[20 160 775 745]{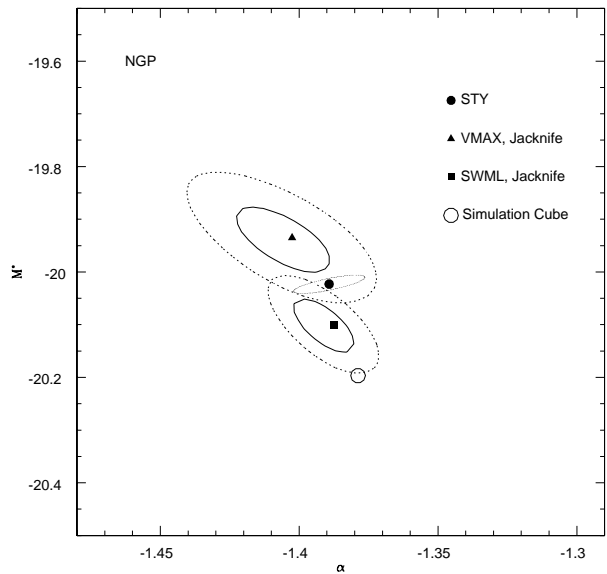}}
\caption{
Contour plots for the best fit luminosity function parameters 
found from the mock catalogue, compared to the underlying
simulation value. Solid lines: $1$-$\sigma$ contours. Dotted lines:
$2$-$\sigma$ contours.
}
\label{fig:levelngp}
\end{figure}

The first check we perform on the mock catalogues is
that we actually recover the Schechter function
parameters of the underlying luminosity function
in the numerical simulation.  Figure \ref{fig:levelngp}
shows the likelihood contours on the $\alpha-M^*$ plane
that correspond to the $1-$ (solid lines), and $2$-$\sigma$ (dotted lines)
levels,
using different methods for estimating the luminosity
function.  The large empty circle represents the
simulation underlying luminosity function, and as can
be seen, the SWML method provides the most accurate
answer from the mock NGP region.

\begin{figure*}
\begin{picture}(450,230)
\put(0,0){\psfig{file=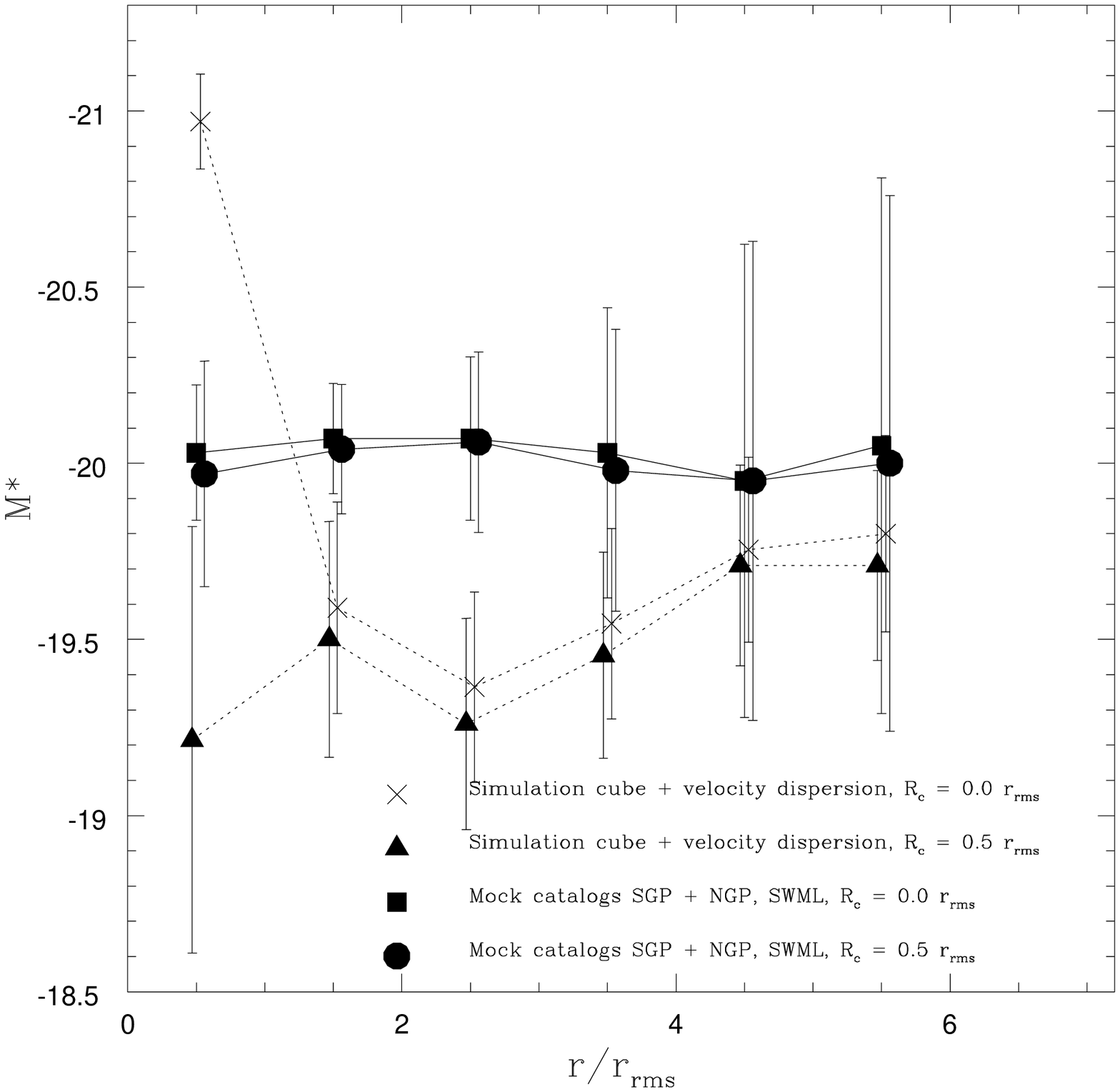,width=8.cm}}
\put(235,0){\psfig{file=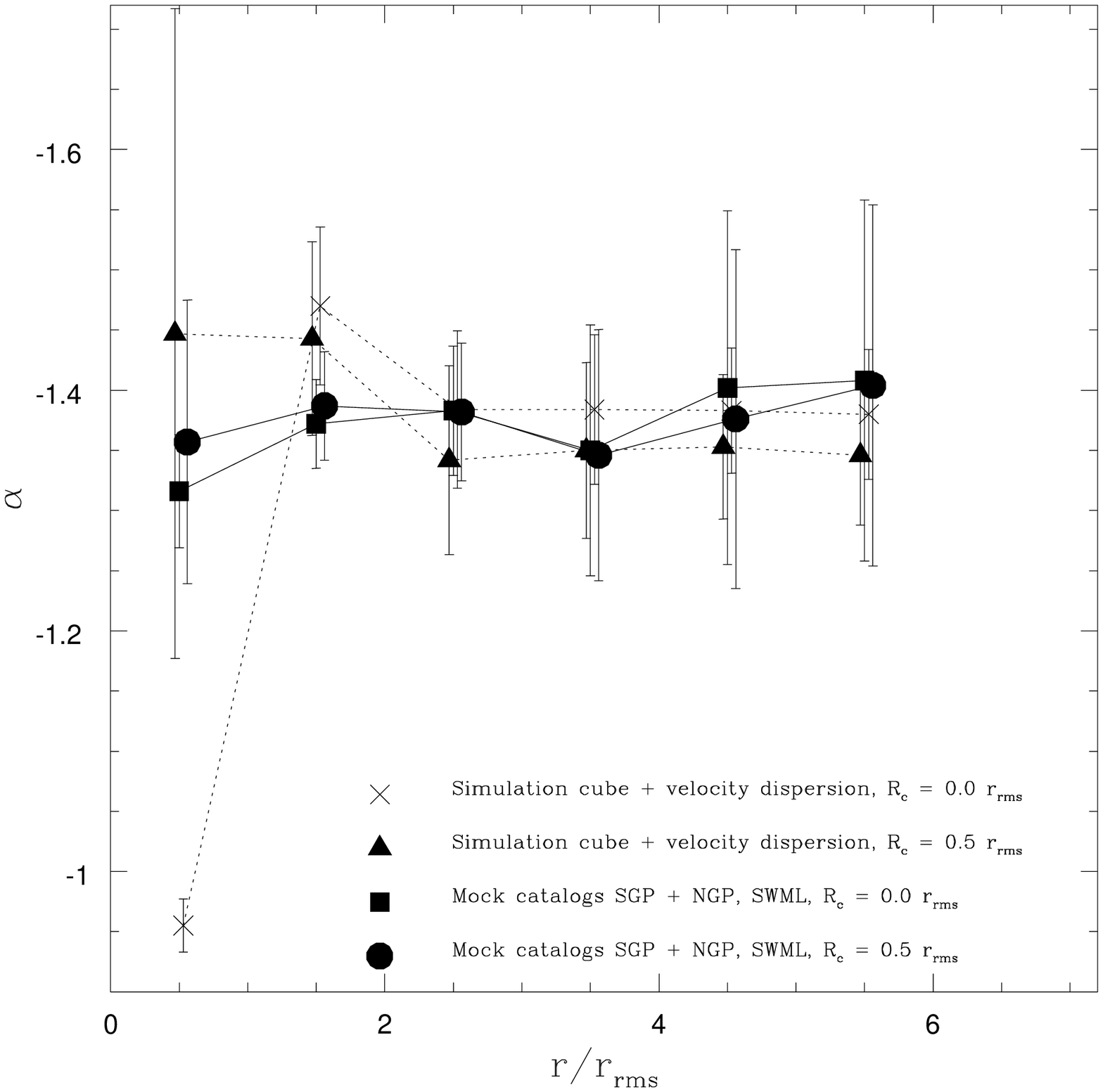,width=8.cm}}
\end{picture}
\caption{
Left panel:
Variation of the characteristic absolute magnitude, $M^*$
as a function of normalised distance
to the group center in the mock catalogues (solid lines).  
For comparison,
results from the full simulation box are also shown (dotted lines). Different symbols
correspond to different choices of cylinder
radii removed to avoid contamination of group inner galaxies on the outer shells,
due to the galaxy velocity dispersion in groups.
Right panel:
Same as top panel but for the  $\alpha$ parameter.
}
\label{fig:parmock}
\end{figure*}

After checking that we can recover the
galaxy luminosity function within at least $2$-$\sigma$ of the underlying values, 
we proceed to rank galaxies
according to their minimum normalised distance to
galaxy groups, just as we did with the semi-analytic
galaxies in the simulation box (in the real and mock catalogues we actually
use the projected positions of group member galaxies to
estimate the group radius, $r_{rms}$).  The resulting luminosity
function parameters can be seen in Figure \ref{fig:parmock}, for
for a cylindrical volume of radius $r_c=0.5r_{rms}$,
that is removed from the analysis in order to avoid the
finger-of-god effects.  For comparison, this Figure also shows the
results from the full numerical simulation box in real-space.
As expected, it is difficult to recover exactly the behaviour
for $\alpha$ and $M^*$ seen in the simulation box.  Moreover,
there is no apparent trend in the Schechter parameters
with $r/r_{rms}$, except for a marginal signature of a minimum
in the $\alpha$ parameter at 3 $r/r_{rms}$, which is
in agreement with the simulation box results.

Still, it is extremely important that
the results are not seriously affected by spurious variations of $M^*$ and $\alpha$,
incompatible with the simulation box results. Therefore, any features in $M^*$ and $\alpha$
seen in the observational data will likely correspond to actual underlying changes in
these parameters.

\section{2dFGRS results}
\label{sec:results}

We start this Section by corroborating that our algorithm provides reliable measurements
of the LF by first calculating the 2dFGRS LF using
the full sample of galaxies and comparing it to previous
measurements.  Applying the SWML algorithm
we find the Schechter parameter combination
 $\alpha=-1.26 \pm 0.03$, $M^*=-19.84 \pm 0.11$,
and $\phi^*=0.0159 \pm 0.0030$.  These results
are in agreement ($\simeq 1$-$\sigma$) with measurements
of the $B_J$ galaxy luminosity function
from the ESO Slice Project by Zucca et al. (1997), who
find $\alpha=1.22\pm0.07$, $M^*=-19.61\pm0.08$ and
$\phi^*=0.020\pm0.004$; and a measurement of the
2dFGRS galaxy luminosity function by Norberg et al. (2002),
who obtain $\alpha=1.21\pm0.02$, $M^*=-19.66\pm0.06$ and
$\phi^*=0.0161\pm0.0005$.  It should be noted that we do not
intend to provide a new measurement of the galaxy luminosity function
in the 2dFGRS, this is only to demonstrate that the code used in this
work provides reasonable answers.

The reader should bear in mind that the semi-analytic model is an 
approach that aims to reproduce the outcome of many real processes involved
in the formation of galaxies, and that this result is tied to a particular 
cosmological model.  Therefore, on a direct comparison between mocks and 
real data,  several characteristics of the numerical and semi-analytical 
models should be taken into account
(For instance, differences in the strength of the peculiar velocity field
between mock and real data can produce different responses
to the method when using real data).  Therefore, the actual variations
of the Schechter parameters as a function of $r/r_{rms}$
may be completely different to what we found in the 
semi-analytic simulation.  We bear these possible differences in mind
and proceed to apply
the method to the real 2dFGRS and 2PIGG catalogues following
the same steps from the previous Section.

\begin{figure*}
\begin{picture}(430,450)
\put(0,0){\psfig{file=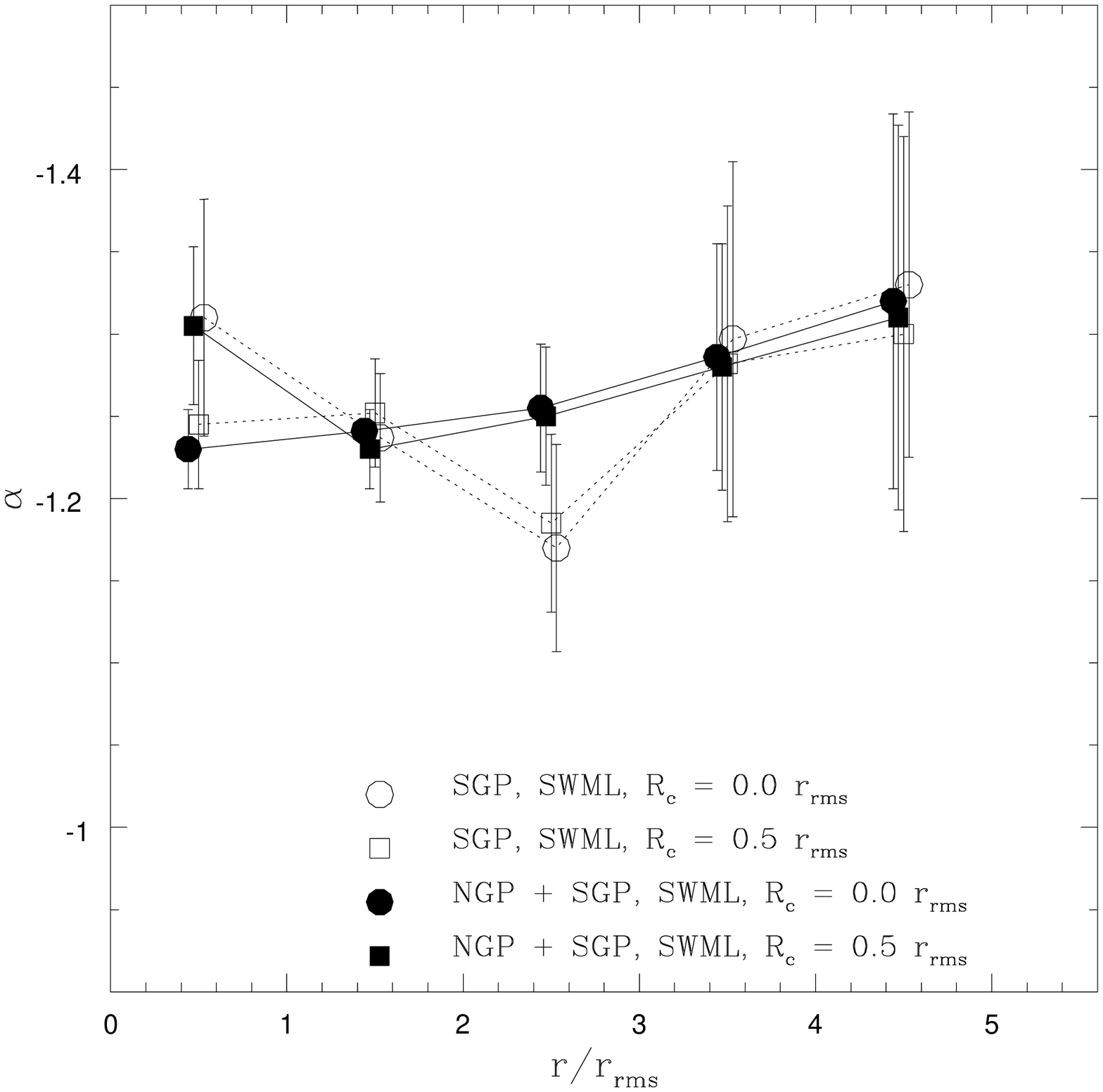,width=8.cm}}
\put(0,215){\psfig{file=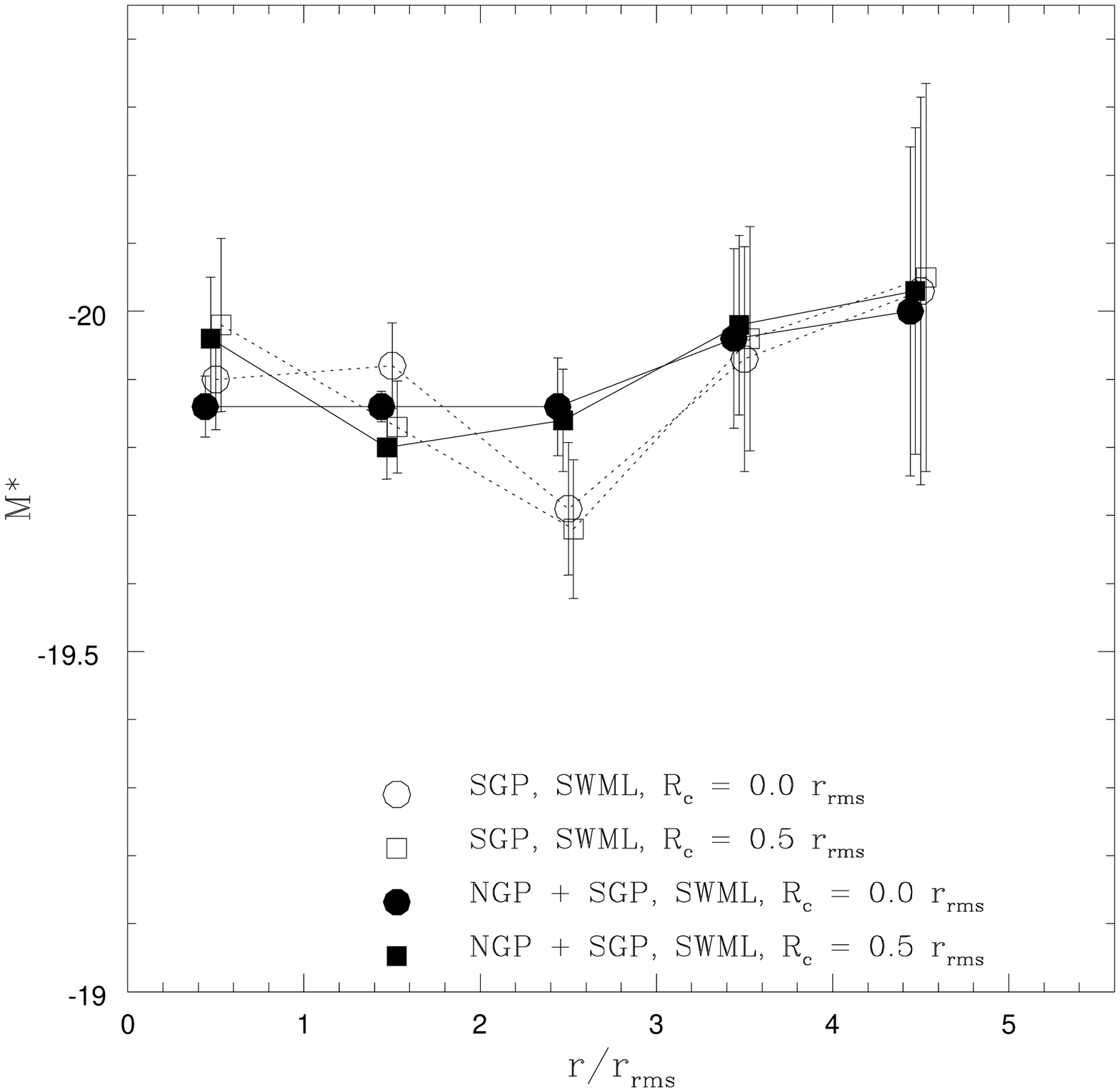,width=8.cm}}
\put(235,0){\psfig{file=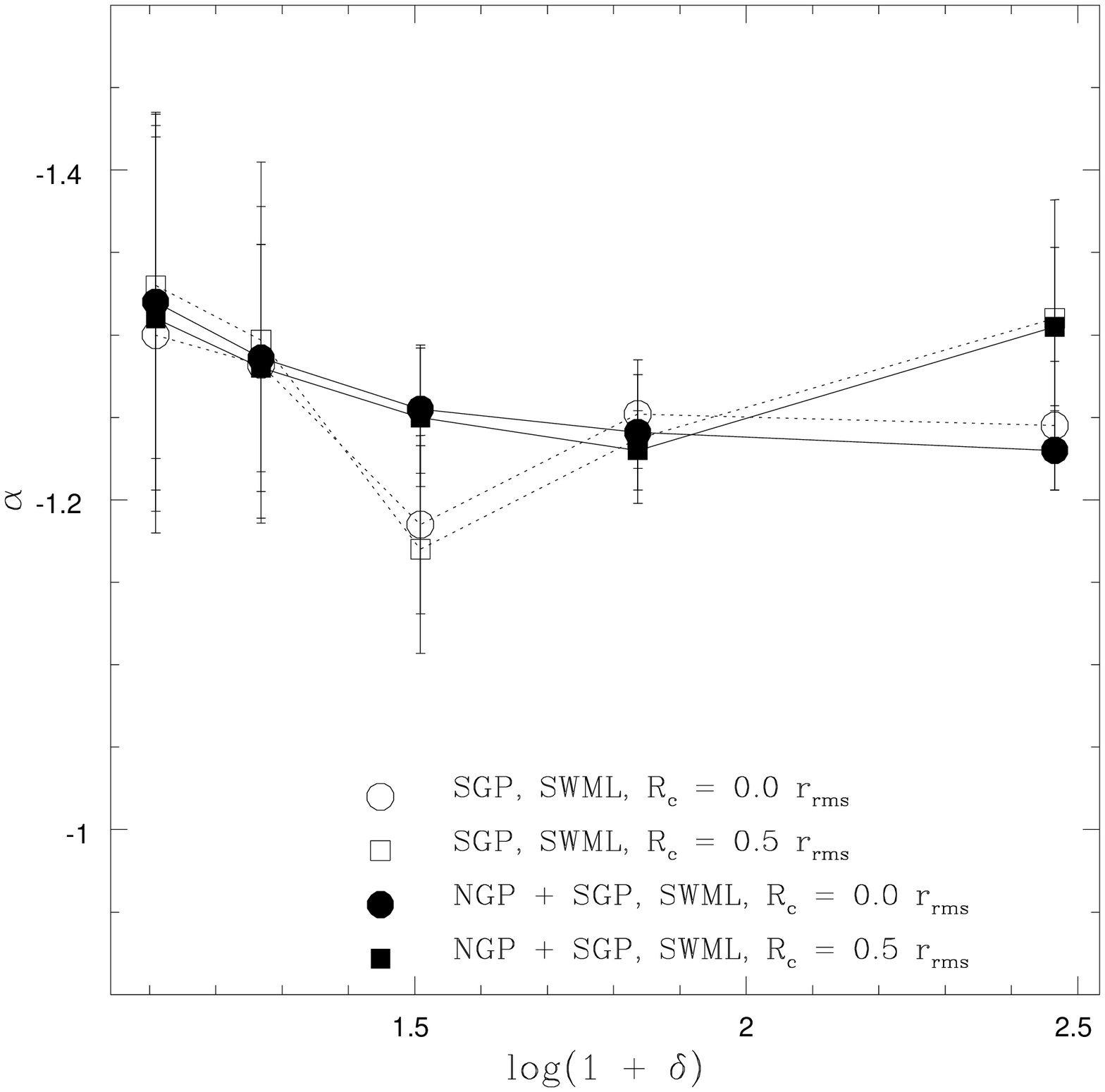,width=8.cm}}
\put(235,215){\psfig{file=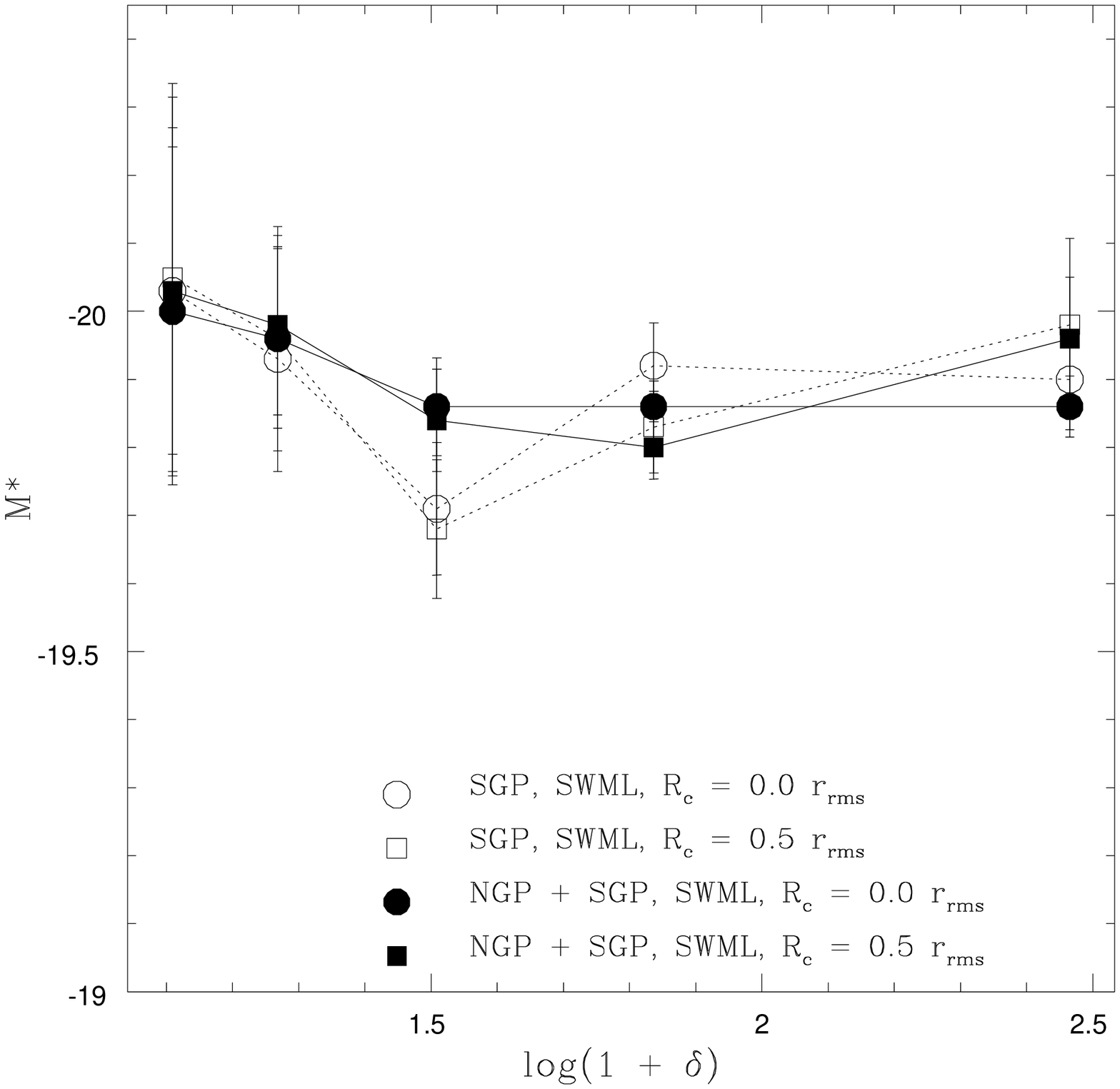,width=8.cm}}
\end{picture}
\caption{
Top panels show the
variation of the characteristic absolute magnitude, $M^*$ as a 
function of normalised distance to the group center (left panels) 
and dark matter density (right panels) 
in the 2dFGRS.  Different lines correspond to different cylinder
radii to avoid contamination of group inner galaxies on outer shells
due to the velocity dispersion of the groups.  Bottom panels are the
same as top panels but for the LF faint-end slope, $\alpha$.
}
\label{fig:par2df}
\end{figure*}

After ranking the 2dFGRS galaxies by $r/r_{rms}$ we 
measure the luminosity functions and present the
Schechter parameters $\alpha$ and $M^*$ in Figure \ref{fig:par2df}.
In this figure, the left panels show the dependence with normalised
distance, and the right panels the dependence on $1+\delta$, which
assuming 
a NFW profile around the 2PIGG galaxy groups,
corresponds to the
dark matter density in terms of the mean density. 

Results from the NGP region are similar to those from the
mock catalogues in that there is no clear trend in the
Schechter parameters with $r/r_{rms}$ given the error bars.  However,
the SGP region shows a dip in both, $M^*$ and $\alpha$
at the expected $r/r_{rms}=3$.
At $r/r_{rms} \simeq 6$ the Schechter parameters converge to values
marginally consistent with the field luminosity function parameters
\cite{norberg,delap}.

This result is very encouraging and may be interpreted
as a possible, new success of the semi-analytic galaxy
formation model.  The implications for this are
dramatic since only one parameter, namely the mass of a
dark-matter halo can reproduce an intuitive
life cycle for a galaxy/halo in the nearby regions of clusters,
including passages, cannibalism, and stripping of dark-matter from tidal
encounters.  Still, the reader should be reminded that
this is only a marginal detection on the SGP region of the
2DFGRS.

The right panels of Figure \ref{fig:par2df} show the
dependence of the Schechter parameters on density contrast
$1+\delta$.  When analysing these results it should be 
kept in mind that the density contrast quoted in these
panels corresponds to that of a NFW profile centred on
the galaxy groups.  Were this profile
a fair representation of the dark-matter density profile
of real galaxy groups, then our values of $1+\delta$
would correspond to the actual dark-matter density
in which our samples of galaxies are embedded.
These results are compatible and complementary
to those from Mo et al. (2004), who study the changes in
$M^*$ and $\alpha$ in the range $-1<\log(1+\delta)<1$. 
There is a clear overlap between our
results and Mo et al. at $\log(1+\delta)=1$, and we
extend this measurement to $\log(1+\delta)=2.5$.

\begin{figure*}
\begin{picture}(450,230)
\put(0,0){\psfig{file=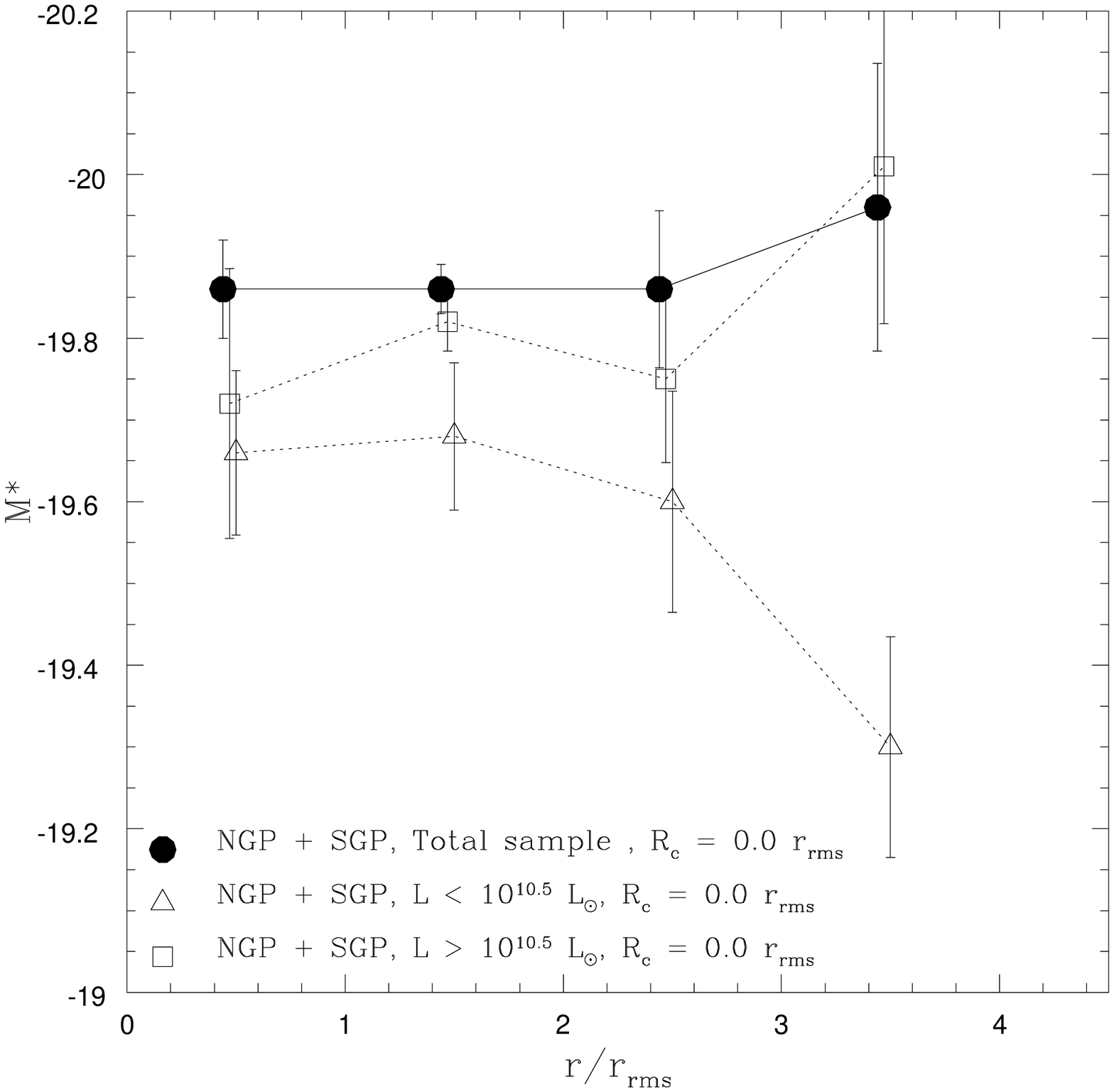,width=8.cm}}
\put(235,0){\psfig{file=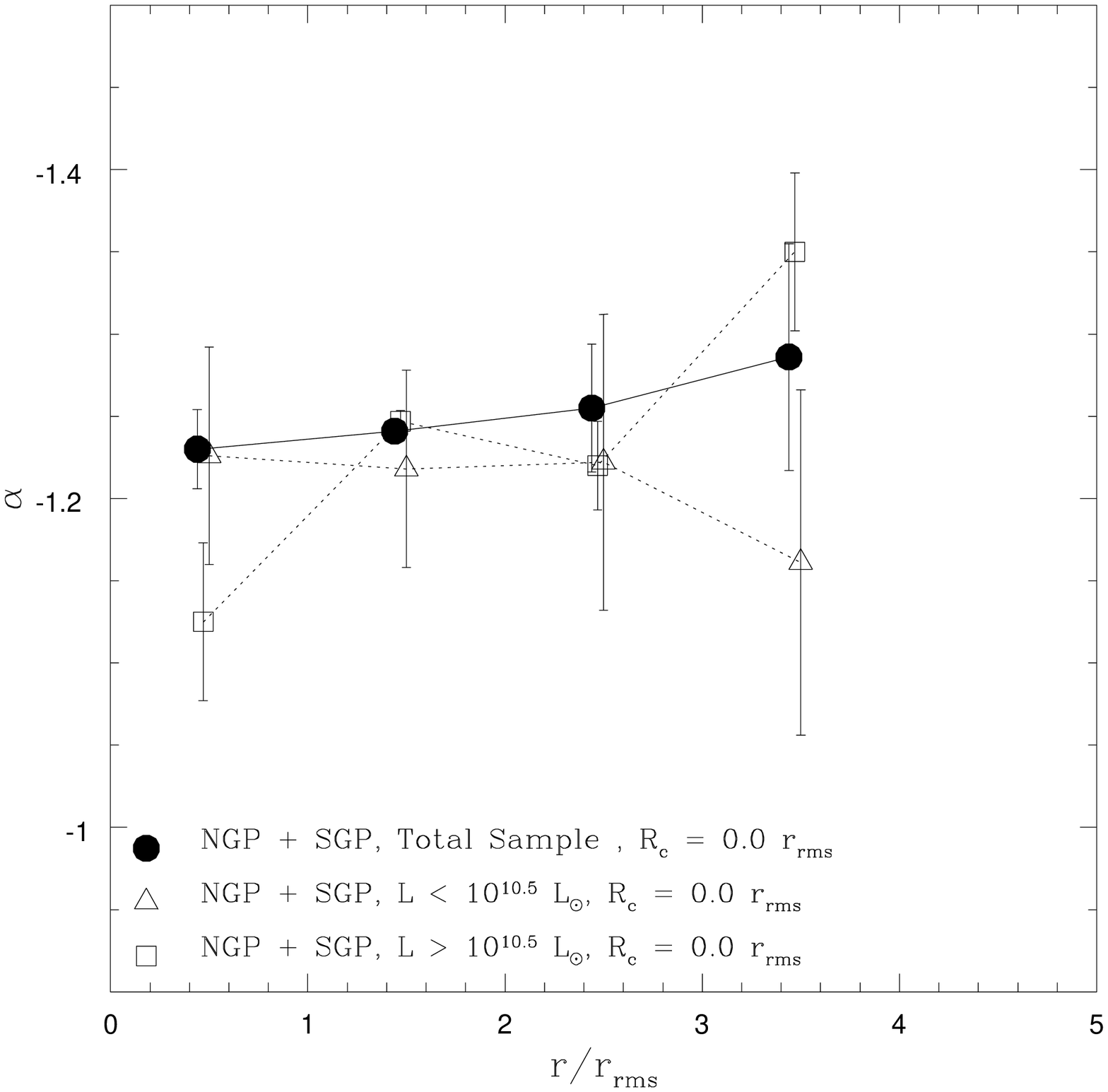,width=8.cm}}
\end{picture}
\caption{
Left panel:
Variation of the characteristic absolute magnitude, $M^*$
as a function of normalised distance
to the group center in the 2dFGRS.  Different lines 
correspond to different galaxy luminosities.
Right panel:
Same as top panel but for the $\alpha$ parameter.
}
\label{fig:2dflum}
\end{figure*}

We perform a final analysis of the luminosity function around
groups. This time we take the full 2dFGRS catalogue and divide 
our sub-samples of galaxies at different $r/r_{rms}$ by
the luminosity of the nearest galaxy group at $L=10^{10.5}L_{\odot}$.
Figure \ref{fig:2dflum} shows the results, and indicates that 
the luminosity function of galaxies around the brighter groups
contain an excess of dwarf galaxies when compared to fainter
groups, at least for $r/r_{rms}>1$, as indicated
by the more negative values of the parameter $\alpha$.
Also, as expected, the typical luminosity of objects
in brighter groups is higher than in fainter groups, as
indicated by the more negative values of $M^*$ in bright
groups.  In this case, this effect is seen at all values
of $r/r_{rms}$. If we compare the LF for different halo masses in Figure \ref{fig:cube}
with the luminosity dependence in Figure \ref{fig:2dflum} we can see 
that brighter groups behave similarly than 
massive haloes except for the inner regions $r/r_{rms}<2$ where
the results in both figures do not match; However, at $r/r_{rms} \simeq 3$ we 
can see the dip of $\alpha$ and $M^*$ at larger normalised distance
for low mass haloes or less luminous haloes.  It has been shown that the total 
luminosity of a group is an excellent tracer of group mass (Eke et al. 2004, 
Padilla et al. 2004), as our results also show.  This Figure also shows
a marginal detection of a more extended dip in $M^*$ for the low Luminous clusters,
although it would be necessary to study a larger sample of groups and galaxies
than the 2dFGRS dataset to be able to measure the upturn in $M^*$ at larger separations.

\section{Conclusions}
\label{sec:conc}

We presented a study on the variations of the luminosity
function of galaxies around clusters of galaxies in numerical
simulations with semi-analytic galaxies.  The aim of this work
was to quantify the extent of the influence of clusters
on their environments and to assess whether such effects
can be reproduced by a semi-analytic model where the main 
parameter determining the population of galaxies are the halo
masses.

In order to do this, we select a population of haloes from the
numerical simulation which are set to be the clusters of our
simulation.  These are characterised by the same distribution
of mass found for the groups in a mock 2PIGG catalogue.  Therefore,
the variations we find in the semi-analytic model can be considered
as predictions which we also test in this work by applying the
method to the 2dFGRS galaxies and 2PIGG groups.

As mentioned above, the main parameter taken into account
by the semi-analytic model when producing galaxies is the
mass of the halo.  Therefore, the variations in the
LF parameters $M^*$ and
$\alpha$ correspond to variations in the mass
function of haloes at different ranges in $r/r_{rms}$, and
when we measure the variations in the luminosity function we 
are actually measuring changes in the mass function with $r/r_{rms}$.
A good agreement between the variations in the simulated and observed LFs 
indicates that the mass of a halo correlates very well with
the different processes a galaxy may have been subject
to during its life.

The semi-analytic galaxy luminosity function shows important
variations with the distance to the centres of clusters 
in the simulation box.  We find that galaxies are brighter in the
cluster inner regions, and that they show a minimum
in characteristic luminosity at intermediate distances,
$r/r_{rms}\simeq 3$.  At larger separations, the characteristic 
luminosity increases and tends to a constant value,
$M^*\simeq-20$.  The minimum in $M^*$ is interpreted
as a measure of the zone of influence of clusters in terms
of their radius.  We also find a similar behaviour
in $\alpha$, which indicates a massive abundance of low
luminosity galaxies (low mass haloes and/or semi-analytic satellites) 
in the cluster central regions,
which decreases to a minimum at intermediate distances,
and increases again at $r/r_{rms}\simeq 5$ to reach
a constant value, somewhat lower in absolute value than for the full
semi-analytic galaxies.  

We interpret these results as clusters being capable of
cannibalising small haloes from nearby regions 
out to $3$ virial radii from their centre, and also of
stripping the mass of normal haloes.  These two
effects would explain the high luminosity central
galaxies surrounded by numerous low luminosity satellites.
The former can be the result of successive mergers of haloes,
whereas the latter can be thought of as 
captured low mass haloes from 
nearby regions, and relatively high mass haloes
from nearby regions which have been tidally stripped off part of their mass.
This would also explain the lack of high and low
luminosity semi-analytic galaxies from the region around $r/r_{rms}\simeq 3$.

By properly taking into account observational biases, we 
checked whether the variations
found in the luminosity function of galaxies around groups
in the simulation box can be measured using available data
from redshift surveys.  First we study the effect that the velocity
dispersion of galaxies in groups induces in our results when
the galaxy positions are measured in redshift space.
When no corrections for this effect are attempted, we find that
the dependence of the Schechter parameters is almost completely
erased.  One way to avoid this is the removal of cylinders centred
on each group, aligned with the line of sight.  The most convenient
value for the projected radius of the cylinder is about $r_c=0.5 r_{rms}$.

Once we demonstrate that it is possible to find variations
in the luminosity function from redshift data (at least
qualitatively), we test whether this can still be detected
using the 2dFGRS galaxies and 2PIGG groups, which are available to
the astronomical community.  The 2dFGRS is affected by a series of
observational biases such as a complicated incompleteness mask and
radial selection function, which can make it even more difficult to
detect the variations in the luminosity function.  Therefore, we produce
mock catalogues which reproduce
as closely as possible all the observational biases
present in the real data.  When we do this, we find that the dependence
of the Schechter parameter on the normalised distance to group centres
is almost completely lost.  This indicates that the removal of a cylindrical
volume around groups is not enough to recover the underlying variations
in the luminosity function, but that the measurements do not return spurious
answers either.

When applying the method to real data, we find that the
luminosity function of galaxies in the 2dFGRS NGP strip
shows no variations in $M^*$ or $\alpha$ with the normalised distance
to the group centres.  However, we do find a minimum value in
$M^*$ and $\alpha$ in the luminosity function of the SGP region,
in agreement with our results from the semi-analytic model.
Further agreement between the data and the semi-analytic model
is found in the correlation between the luminosity of galaxies
and that of their neighbour galaxy group, which is seen in 
the simulation as a correlation between halo mass and the luminosity
of the surrounding galaxies.
Also, the data is marginally consistent with
a dip in $M^*$ for the low 
luminosity 2PIGG groups, again in agreement with the results from the full 
simulation box.
These results strongly indicate
that the interpretation given to the simulation results could actually
have taken place in the formation history of 2dFGRS galaxies.

We have shown in this work that
the variation of the galaxy luminosity function around groups and clusters
of galaxies is a powerful tool for understanding the processes
that galaxies are subject to in the vicinities of clusters.  
It is left to forthcoming efforts to assess whether the variations
in the LF are also accompanied by changes, for instance, in the surface brightness
of galaxies, as this galaxy property is also believed to be influenced
by the presence of a nearby cluster or group of galaxies (or by 
its absence, as in a low surface brightness galaxy).  This as well
as other galaxy properties can be studied to complement the results
obtained in this work.  A future application
of the method presented here that would benefit from improved statistics is the study of
the regions surrounding groups in the Sloan Digital Sky Survey (Abazajian et al. 2004),
which in the very near future will double the number of galaxies available in the
2dFGRS.  This method can also be applied to deep, wide area
surveys (such as the GALEX medium spectroscopic survey, Morrissey et al. 2005)
which will make it possible to detect variations of galaxy properties around groups
with redshift
providing a study of the evolution of the influence of clusters on their
surrounding environments throughout the history of the Universe.

%\begin{table}
%\caption{\small
%{
%Best fit parameters, $A$ and $B$, for the relation between void radius and
%}}
%\begin{tabular}{ccccc}
%\hline
%\hline
%\noalign{\vglue 0.2em}
%Statistic & Real/Redshift-space & Gx./mass & $A$hMpc$^{-1}$ & $B$\\
%\noalign{\vglue 0.2em}
%\hline
%\noalign{\vglue 0.2em}
% $d_{\rm vmin}$ & Real     &  Mass  &  2.30 & 0.80\\
%\noalign{\vglue 0.2em}
%\hline
%\hline
%\end{tabular}\label{table:fits}
%\end{table}

\section*{Acknowledgments}
This work was supported in part by the DAA-ESO grant at PUC,
NDP was supported by a Proyecto Postdoctoral Fondecyt
no. 3040038, GG is supported by Proyecto Regular Fondecyt no. 1040359.
RG was partially supported by Proyecto Regular Fondecyt no. 1040359.
We acknowledge support from the FONDAP Center for Astrophysics.
We have benefited from helpful discussions with D. G. Lambas.  We
thank the Referee for helpful comments and suggestions.

\end{document}